\documentclass[twoside]{article}
\usepackage{Proc_NTSE_14}

\usepackage{bm}
\begin{document}
\thispagestyle{plain}
\publref{myfilename}

\begin{center}
{\Large \bf \strut
 Electromagnetic deuteron form factors in point form of relativistic quantum mechanics
\strut}\\
\vspace{10mm}
{\large \bf
N. A. Khokhlov }
\end{center}

\noindent{
\small \it Komsomolsk-na-Amure State Technical University}

\markboth{
N. A. Khokhlov}
{
Electromagnetic deuteron FFs in PF of RQM}

\begin{abstract}
A study of the electromagnetic structure of the deuteron in the framework of relativistic quantum mechanics is presented. The  deuteron form factors dependencies on the transferred 4-momentum Q
up to 7.5 Fm$^{-1}$ are calculated. We compare results obtained by different realistic
deuteron wave functions stemming from Nijmegen-I, Nijmegen-II, JISP16, CD-Bonn, Paris and
Moscow (with forbidden states) potentials. The nucleon form factors parametrization consistent with the modern experimental analysis was used as the input data.
\\[\baselineskip]
\end{abstract}

\section{Introduction}

In the Born approximation of a one-photon exchange mechanism the elastic $ed$ scattering
observables are directly expressed through the electromagnetic (EM) deuteron form factors (FFs) \cite{Akhiezer,GilmanGross,Garson}.
Therefore this process allows to extract the EM FF dependencies on the transferred 4-momentum $Q$ in the spacelike region.
Relativistic effects may be essential even at low $Q$   \cite{GilmanGross,Garson}.
There are different relativistic models of the deuteron EM FFs \cite{ArenRitz,AdamAren,Tamura,AllenKlinkFormF,LevPace}.

We apply point-form (PF) relativistic quantum
mechanics (RQM) to treat the elastic electron-deuteron scattering in
a Poincar\'{e}-invariant way. Concepts of the RQM
and an exhaustive bibliography are presented in the review by
Keister and Polyzou \cite{RKM_Review}. The PF is one of the three forms of RQM
proposed by Dirac \cite{Dirac}. The other two are the instant form and
front  form. These forms are associated with  different
subgroups  of the Poincar\'{e} group which may be free of interactions. General method for putting interactions in
generators of the Poincar\'{e} group was derived in \cite{BakamjianThomas}.
It was shown that all the forms are unitary equivalent \cite{Sokolov1984}.
Though each form has certain advantages, there are  important simplifying features of
the PF \cite{Klink}. In the PF  all generators of the homogeneous
Lorentz group (space-time rotations) are free of interactions. Therefore only in the PF
the spectator  approximation (SA) preserves its
spectator character in any reference frame \cite{Lev3,Melde,Desplanques}. For an
electromagnetic $NN$ process it means that  the $NN$
interaction does not affect the photon-nucleon interaction, and
therefore the sum of the one-particle EM current
operators is the  sum of the one-particle EM current
operators in any reference frame (r.~f.).
Two equivalent  SAs of EM current operator for composite systems in PF RQM are
derived in Refs. \cite{Klink,Lev3}. The PF SA was applied to calculate EM FFs
 of deuteron, pion, nucleon \cite{AllenKlinkFormF,PFSA1,PFSA2,Wagen,Amghar,CoesterFF2} with reasonable
results.

Present paper is an extension of our previous investigations where we described
elastic  $NN$ scattering up to 3~GeV of laboratory energy
\cite{NNscatMy3},  bremsstrahlung in the $pp$ scattering  $pp\rightarrow pp\gamma$
\cite{ppgammaMy22}, photodisintegration of deuteron $\gamma D\rightarrow np$ \cite{gammaDMy1,gammaDMy2,gammaDMy3} and exclusive electrodisintegration of deuteron  \cite{ednpMy}.
Here we show that the developed approach is applicable to the elastic $eD$ scattering.

\section{Potential model in PF of RQM}
In PF of RQM a system of two particles is described by the
wave function, which is an eigenfunction of the mass operator
$\hat{M}$. We may represent this wave function as a
product of the external and internal parts. The internal
wave function $\vert\chi\rangle$ is also an eigenfunction of the
mass operator and for system of two nucleons with masses
$m_{1}\approx m_{2}\approx m=2m_{1}m_{2}/(m_{1}+m_{2})$ satisfies the equation
%
\begin{equation}
\label{rel1} \hat{M}\vert\chi\rangle\equiv \left[ 2\sqrt{{{\bf
q}}^2+m^2}+V_{int} \right]\vert\chi\rangle = M\vert\chi\rangle,
\end{equation}
where $V_{int}$ is  an operator commuting with the full angular
momentum operator and acting only through internal variables (spins
and relative momentum), ${{\bf q}}$ is a momentum operator of one of
the particles in the center of mass frame (relative momentum).
Rearrangement of (\ref{rel1}) gives
\begin{equation}
\label{rel2} \left[ {{{\bf q}}^2 + m V} \right]\vert\chi\rangle = q^2\vert\chi\rangle,
\end{equation}
where $V$ acts like $V_{int}$ only through internal variables and
\begin{equation}
  q^2 = \frac{M^2}{4} - m^2.\label{q2}
\end{equation}
Eq.~(\ref{rel2}) is identical in form to the Schr\^{o}dinger
equation. The only relativistic correction here is in the deuteron binding energy
that must be changed  by the effective value $2.2233$ MeV instead of the experimental $2.2246$ MeV.
It is easy to show. Let $\varepsilon$ be the deuteron binding energy.
Then $M=2m-\varepsilon$ and for deuteron state $q^2=\frac{M^2}{4}-m^2=-m\varepsilon \left(1-\frac{\varepsilon}{4m}\right)$.
Comparing with the nonrelativistic relationship $q^2=-m\varepsilon$
we can identify factor $\left(1-\frac{\varepsilon}{4m}\right)$ as the essential relativistic correction.
There is no similar correction in the scattering region because $q^2=mE_{lab}/2$ is the precise relativistic relationship ($E_{lab}$ is the laboratory energy), that is used in the partial wave analysis.
The change is negligible for our problem.
Therefore this formal identity allows us to use non relativistic deuteron wave functions in our calculations.


\section{$eD$ elastic scattering}
We give only results of PF RQM  necessary
for our calculation, in notation of Ref.~\cite{Lev3}. We use
formalism of \cite{Lev3} for calculation of the matrix elements of
the EM current operator.

We consider the $pn$ system and neglect difference of neutron and
proton masses. Let $p_i$ be the 4-momentum  of nucleon
$i$, $P\equiv(P^0,{\bf P})=p_1+p_2$ be the system 4-momentum, $M$ be
the system mass and $G=P/M$ be the system 4-velocity. The wave
function of two particles with 4-momentum $P$ is expressed through a
tensor product of external and internal parts
\begin{equation}
\vert P,\chi\rangle
=U_{12}\,\vert P\rangle\otimes\vert \chi\rangle,
\end{equation}
where the internal wave function $\vert\chi\rangle$ satisfies
Eqs.~(\ref{rel1})-(\ref{rel2}). The unitary operator
\begin{equation}
U_{12}=U_{12}(G,{\bf q})=\prod_{i=1}^{2} D[{\bf
s}_i;\alpha(p_i/m)^{-1}\alpha(G)\alpha(q_i/m)] \label{U12}
\end{equation}
is the  operator from the ''internal'' Hilbert space  to the
Hilbert representation space of two-particle states \cite{Lev3}.
$D[{\bf s};u]$ is the SU(2) representation operator
corresponding to the element $u\in$SU(2).  ${\bf s}$ are generators of the  representation.
In our case of spin $s$ = 1/2 particles, we deal with the
fundamental representation, i.e. ${\bf s}_i\equiv
\frac{1}{2}{\bm \sigma}_i$ (${\bm \sigma} =( \sigma_x,
\sigma_y, \sigma_z)$ are the Pauli matrices) and $ 
D[ {\bf s}; u]\equiv u$.
The momenta of the particles in their c.m. frame are
\begin{equation}
q_i=L[\alpha(G)]^{-1}p_i, \label{scm_q}
\end{equation}
where $L[\alpha(G)]$ is the Lorentz transformation to the frame
moving with 4-velocity $G$. Matrix $\alpha (g)= {(g^0+1+ {\bf  \sigma }\cdot {\bf g)}}/{
\sqrt{2(g^0+1)}}$  corresponds to a 4-velocity $g$.

The external part of the wave function is defined as
\begin{gather}
\label{ext_wf}  \langle G\vert
P'\rangle\equiv\frac{2}{M'}G^{'0}\delta^3({\bf G}-{\bf G}').
\end{gather}
Its scalar product is
\begin{gather}
\label{ext_wf_product}  \langle P''\vert P'\rangle=\int
\frac{d^3{\bf G}}{2G^0} \langle P''\vert G\rangle \langle G\vert
P'\rangle=2\sqrt{M^{'2}+{\bf P}^{'2}}\delta^3({\bf P''}-{\bf P}'),
\end{gather}
where $G^{0}({\bf G})\equiv\sqrt{1+{\bf G}^2}$.
The internal part
 $\vert\chi\rangle$ is characterized by momentum  ${\bf q}={\bf q}_1=-{\bf q}_2$ of one of the particles in the
 c.m. frame.

 According to the Bakamjian---Thomas procedure \cite{BakamjianThomas}
interaction appears in 4-momentum $\hat{P}=\hat{G}\hat{M}$, where $\hat{M}$ is sum of the free mass
operator ${M}_{free}$ and of the interaction $V_{int}$: $\hat{M}={M_{free}}+V_{int}$
(Eq.~(\ref{rel1})). The interaction operator acts only
through internal variables. Operators  $V_{int}$
and $V$ (and therefore  $\hat{M}$ and  ${M}_{free}$) commute with spin operator $S$ (full angular momentum) and
with 4-velocity operator $\hat{G}$. Generators of space-time rotations  are free of interaction.
Most non-relativistic scattering theory  formal results are valid
for  case of two particles \cite{RKM_Review}. For example in the c.m. frame
the relative orbital angular momentum and spins are
 coupled together as in the non-relativistic case.

The deuteron wave function $\vert P_{i},\chi_{i}\rangle$ is
normalized as follows
\begin{equation}
\label{sc_prod_deut} \langle P_f,\chi_{f}\vert
P_i,\chi_{i}\rangle=2 P_{i}^{0}\,\delta^3({\bf P}_i-{\bf
P}_f)\langle \chi_{f}\vert
\chi_{i}\rangle.
\end{equation}


   There is a convenient r.~f. for calculation of current operator matrix elements \cite{Lev3} (it coincides for  elastic $ed$ scattering with the Breit r.~f.).
   For all EM reactions with two nucleons this r.~f. is defined  by condition:
\begin{equation}
{\bf G}_{f}+{\bf G}_{i}=0,
 \label{cond1}
\end{equation}
where ${\bf G}_f={\bf P}_f/M_D$, ${\bf G}_i={\bf P}_i/M_D$ are final and initial 4-velocities of the deuteron and $M_D$ is its mass.
 The matrix element of the current operator is  \cite{Lev3}:
\begin{equation}
\langle P_{f},\chi_{f}\vert \hat{J}^{\mu}(x)\vert P_{i},\chi_{i}\rangle=2(M_{f}M_{i})^{1/2} \exp(\imath (P_f - P_i)x)\langle\chi_f\vert \hat{j}^{\mu}({\bf h})\vert\chi_i\rangle , \label{mat1}
\end{equation}
where $\hat{j}^{\mu}({\bf h})$ defines action of current operator in the internal space of the $NN$ system.

\begin{equation}
{\bf h}=\frac{2(M_i M_f)^{1/2}}{(M_i+M_f)^{2}}\,{\bf k}=\frac{\bf k}{2M_{D}}
\end{equation}
is vector-parameter \cite{Lev3} ($0\leq h \leq 1$), ${\bf k}$ is momentum of photon in r.~f.  (\ref{cond1}), $M_{i}=M_{f}=M_{D}$ are masses of initial and final $NN$ system (deuteron).

The internal wave function of deuteron  is
\begin{gather}
\vert\chi_i\rangle=
\frac{1}{r}\sum_{l=0,2}u_{l}(r)\vert l,1;J=1M_{J} \rangle_{{\bf r}} \label{wf_D}
\end{gather}
with normalisation  $\langle \chi_i\vert\chi_i\rangle=1$.
We use the momentum space wave function:
\begin{gather}
\vert\chi_i\rangle=\frac{1}{q}\sum_{l=0,2}u_{l}(q)\vert l,1;1M_{J} \rangle_{{\bf q}},
\end{gather}
where
\begin{align}
u(q)&\equiv u_{0}(q)=\sqrt{\frac{2}{\pi}}\int dr \sin(qr) u(r), \\
w(q)&\equiv u_{2}(q)=\sqrt{\frac{2}{\pi}}\int dr \left[ \left(\frac{3}{(qr)^2}-1\right)\sin(qr)-\frac{3}{qr}\cos(qr)\right]w(r).
\end{align}

Transformations from the Breit r.~f. (\ref{cond1}) to the initial c.~m. frame of the $NN$ system and  to the final one are boosts along vector
 ${\bf h}$ (axis $z$). Projection of the total deuteron angular  momentum onto the  $z$ axis does not change for these boosts.
The initial deuteron in the Breit r.~f.  moves in direction opposite  to the ${\bf h}$. Its internal  wave function with spirality   $\Lambda_i$
  is
\begin{equation}
\vert \Lambda_i\rangle=\frac{1}{q}\sum_{l=0,2}u_{l}(q)\vert
l,1;1,M_{J}=-\Lambda_i \rangle. \label{wi_D_spir}
\end{equation}
 Wave function of the final deuteron with spirality   $\Lambda_f$ is
\begin{equation}
\vert \Lambda_f\rangle =\frac{1}{q}\sum_{l=0,2}u_{l}(q)\vert
l,1;1,M_{J}=\Lambda_f \rangle, \label{wf_D_spir}
\end{equation}

Usual parametrisation of the EM CO matrix element for the deuteron (spin 1 particle) is  \cite{Arnold81,GilmanGross,Garson}:
\begin{multline}
(4P_i^{0}P_f^{0})^{1/2}\langle P_f,\chi_f|J^{\mu} |P_i,\chi_i\rangle \\
=- \left\{ G_1(Q^2)({\bm\xi}_f^{*}\cdot {\bm\xi}_i)-G_3(Q^2)\frac{({\bm\xi}_f^{*}\cdot \Delta P)({\bm\xi}_i\cdot \Delta P)}{2M_{D}^2}\right\}(P_i^{\mu}+P_f^{\mu}) \\
-G_2(Q^2)[{\xi}_i^{\mu}({\bm\xi}_f^{*}\cdot \Delta\textbf{ P})-{\xi}_f^{*\mu}({\bm\xi}_i\cdot \Delta \textbf{P})],
\end{multline}
where $(a\cdot b)=a^0 b^0 -({\bf a}\cdot {\bf b})$,  form factors $G_i(Q^2)$, $i=1,2,3$ are function of $Q^2=-\Delta P^2$, $\Delta P=P_f-P_i$.

In the Breit r.~f. $\textbf{P}_f=-\textbf{P}_i$,  $P_i^{0}=P_f^{0}\equiv P^0=M_D /\sqrt{1-h^2}$, ${\Delta P=(0, 2\textbf{P}_f)}$,
${P_i^{\mu}+P_f^{\mu}=(2P^0,\textbf{0})}$, ${\bf P}_f/P^0={\bf h}$, ${\bf P}_f={\bf h}M_D/\sqrt{1-h^2}$, ${\Delta P^2=-4h^{2} M_D^2/(1-h^2)}$, ${Q^2\equiv -\Delta P^2}$,
$h^2=({\bf h}\cdot {\bf h})$.

\begin{multline}
\langle \chi_f|j^{0}({\bf h}) |\chi_i\rangle =
 - G_1(Q^2)({\bm\xi}'^{*}\cdot {\bm\xi})+2G_3(Q^2)\frac{({\bm\xi}_{f}^{*}\cdot {\textbf{ h}})({\bm\xi}_i\cdot {\textbf{h}})}{1-h^2}\\
+G_2(Q^2)[{\xi}_i^{0}({\bm\xi}_{f}^{*}\cdot {\bf h})-{\xi}_f^{0*}({\bm\xi}_i\cdot  \textbf{h})],
\end{multline}
\begin{equation}
\langle \chi_f|{\bf j}({\bf h}) |\chi_i\rangle =
G_2(Q^2)[\xi_i({\bm\xi}_f^{*}\cdot {\bf h})-\xi_f^{*}({\bm\xi}_i\cdot {\bf h})]=
G_2(Q^2)[{\bf h}\times [{\bm\xi}_i\times {\bm\xi}_f^{*}]].
\end{equation}

It can be shown  \cite{Lev3}, that these expressions are equivalent to choosing  $j^{\nu}$ as:
\begin{align}
j^0({\bf h})&=
G_C(Q^2)+\frac{2}{(1-{h}^2)}G_Q(Q^2)
\left[\frac{2}{3}{ h}^2-({\bf h}\cdot {\bf J})^2\right],\label{123}\\
{\bf j}({\bf h})&=-\frac{\imath }{\sqrt{1-{h}^2}}
G_M(Q^2)({\bf h}\times {\bf J}),
\end{align}
where ${\bf J}$ is total angular momentum (spin) of the deuteron. $G_C$, $G_Q$, $G_M$ are its charge monopole, charge quadruple and magnetic dipole FFs.

Spiral polarizations  of the deuteron in the initial and final states are
\begin{align}
\xi_{i}^{\Lambda}&=\left\{
\begin{array}{l@{{}\ \ \ \ \ \ \ \ {}}r}
(0,\pm 1,-\imath,0)/\sqrt{2}& (\Lambda=\pm) \\
(-Q/2,0,0,P_{0})/M_D  =(-h,0,0,1)/\sqrt{1-h^2}& (\Lambda=0),\\
\end{array}
\right.\\
 \xi_{f}^{\Lambda}&=\left\{
\begin{array}{l@{{}\ \ \ \ \ \ \ \ \ \ \ \ \ {}}r}
(0,\mp 1,-\imath,0)/\sqrt{2}& (\Lambda=\pm) \\
(Q/2,0,0,P_{0})/M_D = (h,0,0,1)/\sqrt{1-h^2}&(\Lambda=0).\\
\end{array}
\right.
\end{align}

Polarization of the virtual photon is
\begin{equation}
\epsilon^{\lambda}=\left\{
\begin{array}{l@{{}\ \ \ \ \ \ \ \ {}}r}
(0,\mp 1,-\imath,0)/\sqrt{2}& (\lambda=\pm) \\
(1,0,0,0)& (\lambda=0).\\
\end{array}
\right.
\end{equation}

FFs $G_i$ are expressed as
\begin{equation}
\begin{split}
G_C&=G_1+\frac{2}{3}\eta G_{Q},\\
G_{Q}&=G_1-G_M+(1+\eta)G_3,\\
G_1&=G_C-\frac{2h^2}{3(1-h^2)}G_Q,\\
G_3&=G_Q\left(1-\frac{h^2}{3}\right)-G_C(1-h^2)+G_M(1-h^2),
\end{split}
\end{equation}
where $\eta=Q^2/4M_D^2=h^2/(1-h^2)$.

At $Q^2=0$ we have $G_C=G_1$, $G_Q=G_1-G_M+G_3$. Form factors ${G_C(0)=e}$, $G_M(0)=\mu_{D}e/2 M_D$ and $G_Q=Q_{D}e/M^2_D$ give charge, magnetic and quadruple momenta of deuteron.

Denoting helicity amplitudes as  $j^{\lambda}_{\Lambda_{f}\Lambda_{i}} \equiv \langle \Lambda_f | \left( \epsilon^{\lambda}_{\mu}\cdot j^{\mu}({\bf h}) \right) | \Lambda_i \rangle$,
we arrive at:
\begin{equation}
j^{0}_{00}(Q^2)=G_{C}+\frac{4}{3}\frac{h^2}{1-h^2}G_{Q},
\end{equation}
\begin{equation}
j^{0}_{+-}(Q^2)=j^{0}_{-+}(Q^2)=G_{C}-\frac{2}{3}\frac{h^2}{1-h^2}G_{Q},
\end{equation}
\begin{equation}
\frac{j^{+}_{+0}(Q^2)+j^{+}_{0-}(Q^2)}{2}=-\frac{h}{\sqrt{1-h^2}}G_{M}
\end{equation}
and
\begin{equation}
j^{+}_{+0}(Q^2)=j^{-}_{-0}(Q^2)\approx j^{+}_{0-}(Q^2)=j^{-}_{0+}(Q^2).\label{j000}
\end{equation}

 The deuteron FFs are associated with unpolarised structure functions  \cite{Donnelly}:
\begin{gather}
A(Q^2)=G_{C}^{2}(Q^2)+\frac{2}{3}\eta G_{M}^{2}(Q^2),\\
B(Q^2)=\frac{4}{3}\eta (1+\eta)G_{M}^{2}(Q^2).
\end{gather}
These quantities are extracted from the elastic $eD$ scattering with unpolarised particles.
A tensor polarisation observable  $t_{20}(Q^{2},\theta)$ is usually used as an additional quantity needed for definition of all three FFs.

In the present paper we use SA of the EM CO of \cite{Lev3} without expanding it in powers of $h$ and we calculate matrix elements in the momentum space.
Therefore calculating
 (\ref{j000}) we use a following expansion of  $\hat{j}^{\mu}({\bf h})\approx \hat{j}_{SA}^{\mu}({\bf h})$ \cite{ednpMy}
\begin{multline}
\hat{j}_{SA}^{\mu}({\bf h})=\left(1+({\bf A}_{2}\cdot{\bf s}_2)\right)\left(B^\mu_{1}+({\bf C}^\mu_{1}\cdot{\bf s}_1)\right){\bf I}_{1}({\bf h})\\
+ \left(1+({\bf A}_{1}\cdot{\bf s}_1)\right)\left(B^\mu_{2}+({\bf C}^\mu_{2}\cdot{\bf s}_2)\right){\bf I}_{2}({\bf h}),
\end{multline}
where  ${\bf A}_{i}$, $B^\mu_{i}$, ${\bf C}^\mu_{i}$ are some vector functions of  ${\bf h}$ and ${\bf q}(q,\theta,\phi)$.
In the spherical coordinate system $(q,\theta,\phi)$ dependence of these functions on $\phi$ appears as $e^{\pm im\phi}$ ($m=0,1,2$). The $\phi$ is analytically integrated giving trivial equalities in (\ref{j000}).
\section{Results}
\begin{figure}[th]
\centerline{\includegraphics[width=0.8\textwidth]{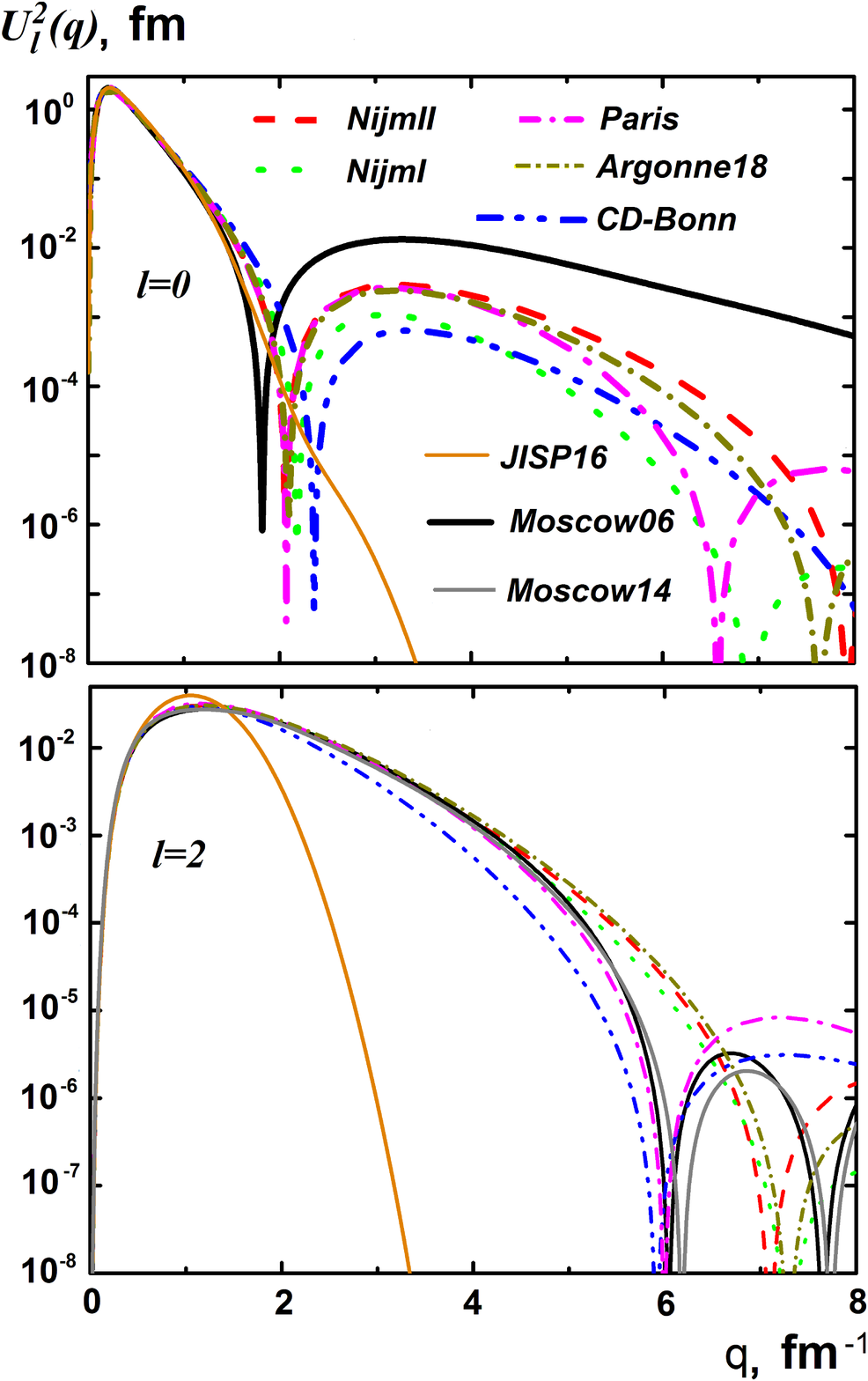}}
\caption{Momentum space deuteron wave functions  used in the calculations. Same legend for $S$- and $D$-wave functions.}
\label{fig1}
\end{figure}

\begin{figure}[ht]
\centerline{\includegraphics[width=0.8\textwidth]{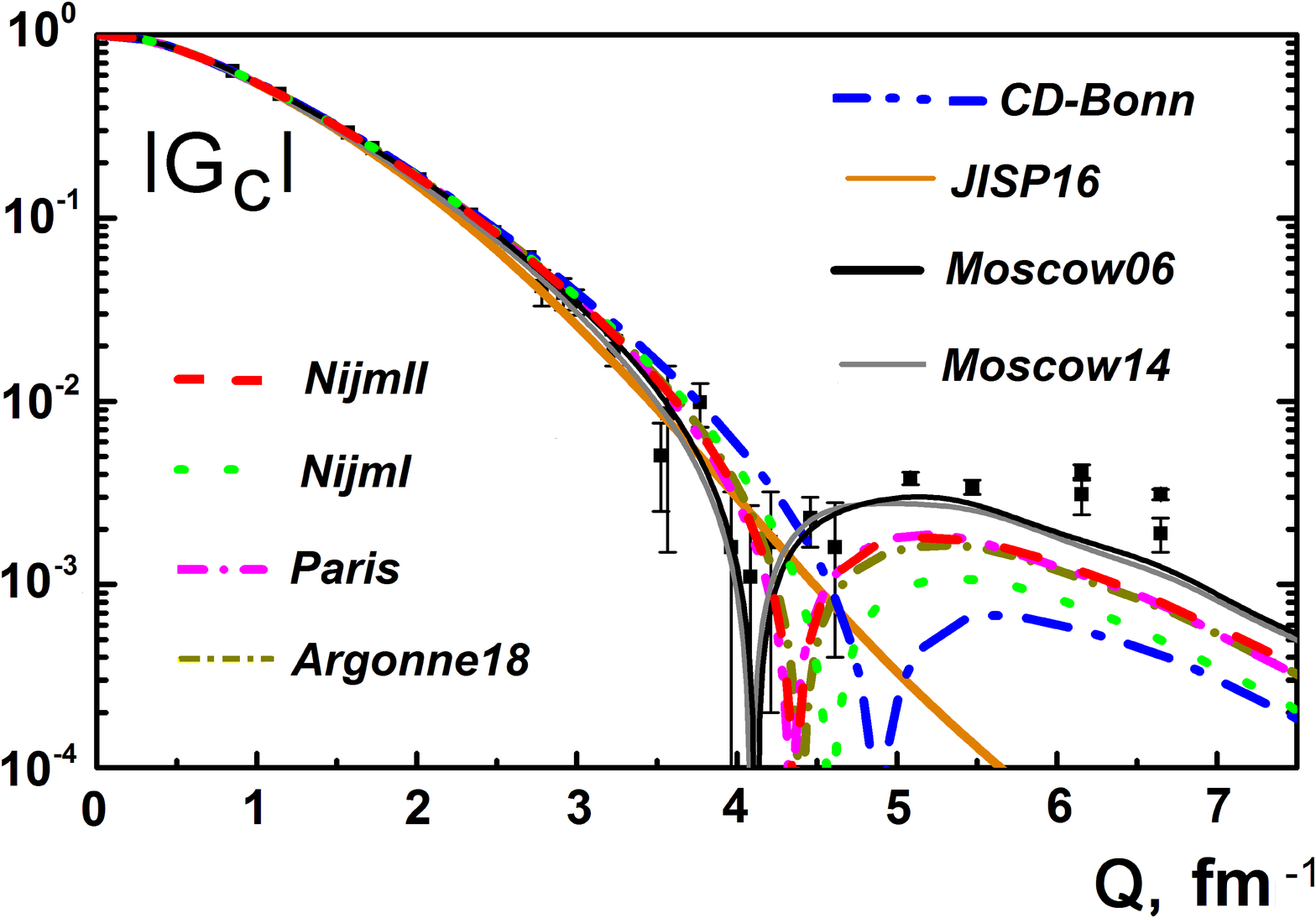}}
\caption{Deuteron form factor $G_{C}$ as a function of Q. Data are from compilation of \cite{Garson}
calculated from $A$, $B$ and $t_{20}$ data of
\cite{r87,r90,r94,r97,r98,r99,r95,r96,r76,r77,r103,r104,r105,r106,r107,r108,r78}.}
\label{fig4}
\end{figure}

\begin{figure}[ht]
\centerline{\includegraphics[width=0.8\textwidth]{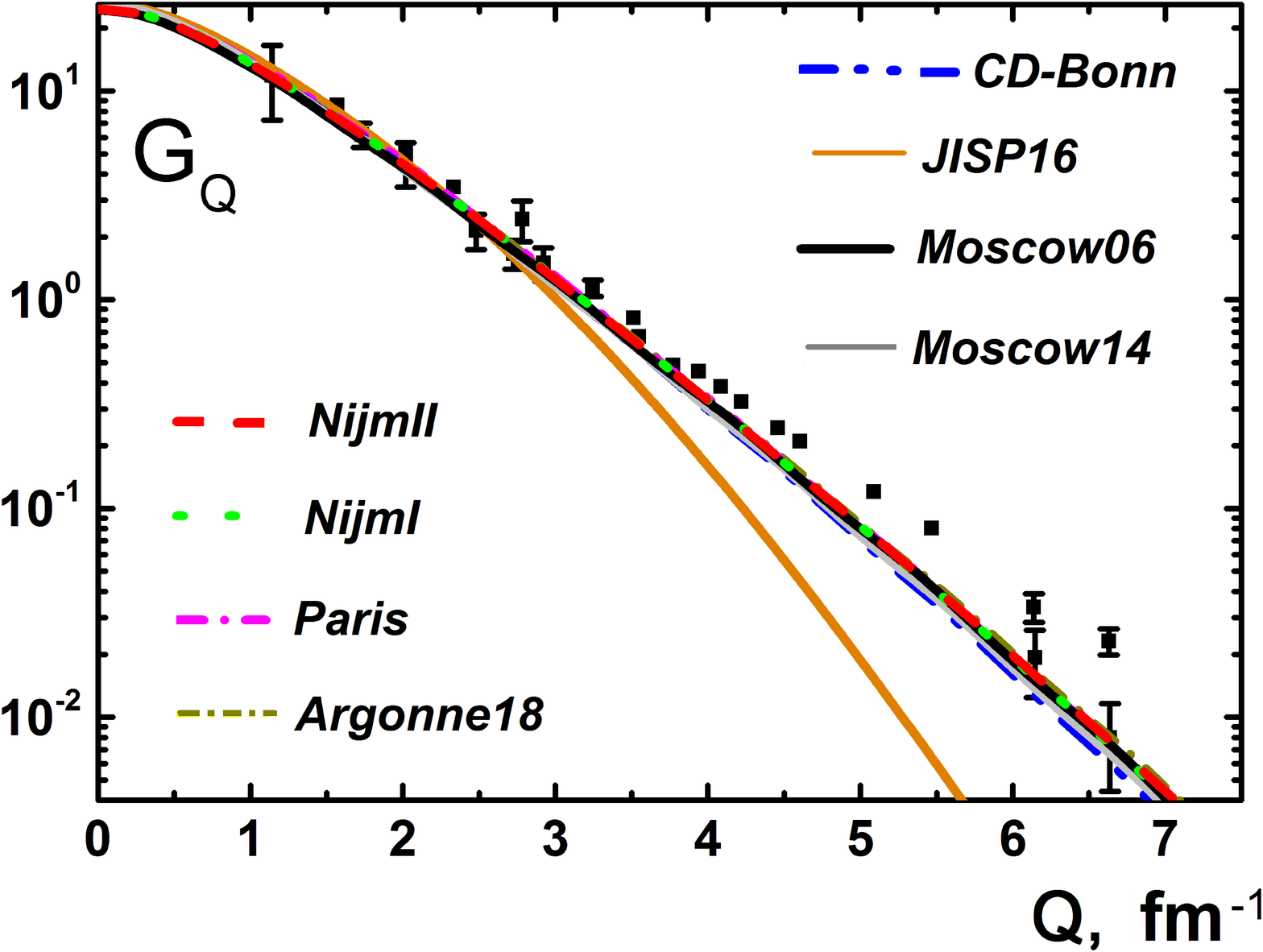}}
\caption{Deuteron form factor $G_{Q}$ as a function of Q. Same legend as Fig.~\ref{fig4}.}
\label{fig5}
\end{figure}

\begin{figure}[ht]
\centerline{\includegraphics[width=0.8\textwidth]{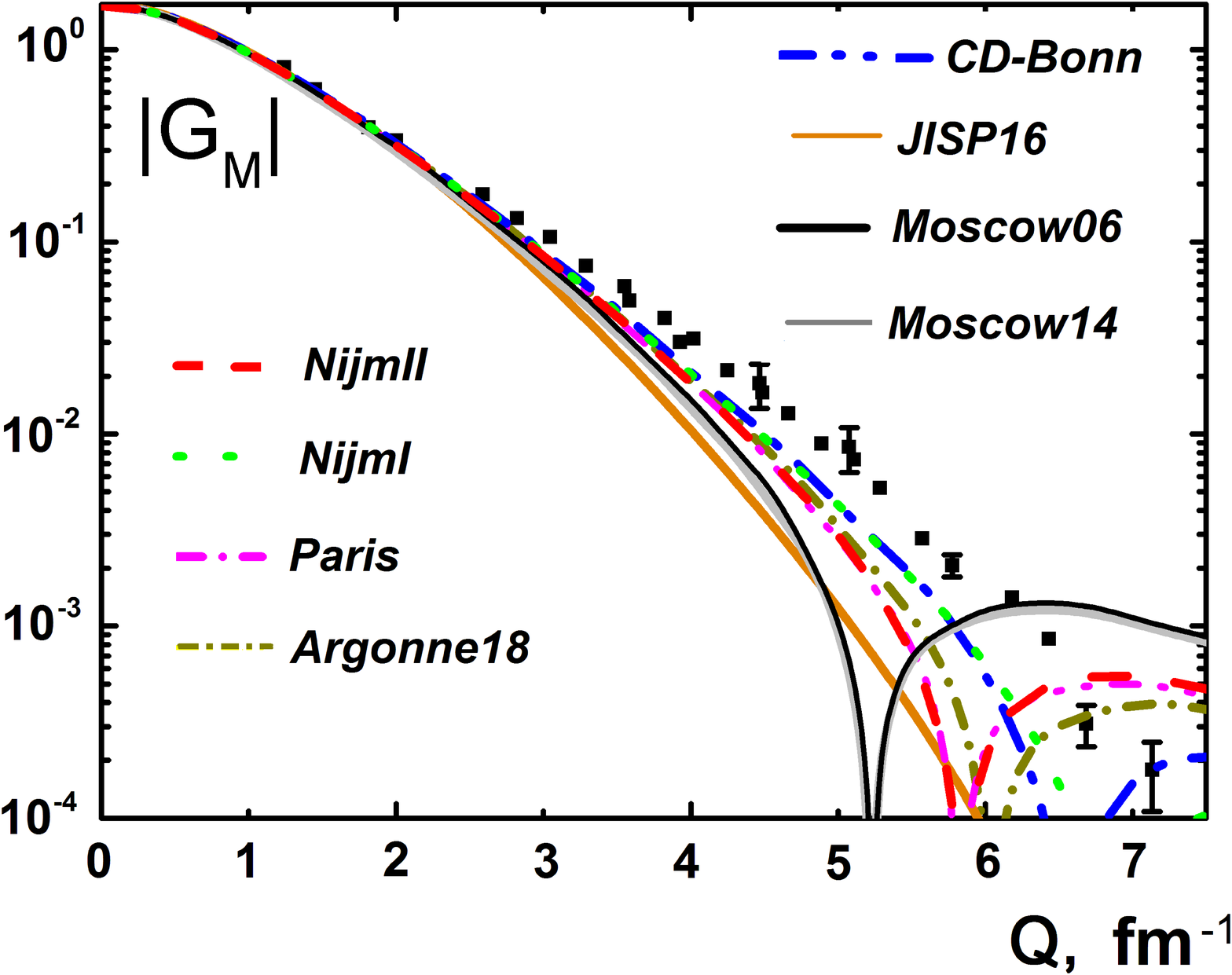}}
\caption{Deuteron form factors $G_{M}$ as a function of Q. Same legend as Fig.~\ref{fig4}}
\label{fig6}
\end{figure}

In our calculation we use as an input the momentum space deuteron wave functions and nucleon EM FFs. The momentum space deuteron wave functions stemming from Nijmegen-I (NijmI), Nijmegen-I (NijmII) \cite{Nijm}, JISP16 \cite{JISP16}, CD-Bonn \cite{CD-Bonn}, Paris \cite{Paris}, Argonne18 \cite{Argonne18} (momentum space deuteron wave function is from \cite{Argonne18ms}) and
Moscow (with forbidden states) \cite{NNscatMy3} potentials are shown in Figs.~\ref{fig1}. We use two versions of Moscow type potential: Moscow06  \cite{NNscatMy3} and Moscow14. Last one was derived by author in the same
manner as in  \cite{NNscatMy3} but with deuteron asymptotic constants fitted to describe static deuteron form factors. Parameters of both Moscow potentials may be requested from the author (e-mail: nikolakhokhlov@yandex.ru).
For all but JISP16  the $S$-wave functions  change sign at $q\approx 2$~Fm$^{-1}$,
and $D$-wave functions change sign at $q\approx 6-8$~Fm$^{-1}$. The $S$- and $D$-wave functions of Argonne18, Paris and  NijmII
are close for $q\lesssim 5$~Fm$^{-1}$. The $S$-wave functions of CD-Bonn and NijmI are close for $q\lesssim 5$~Fm$^{-1}$. Wave functions of the JISP16 decrease rapidly after about 2 Fm$^{-1}$  without changing sign.
Our  results for deuteron EM FFs are presented in Table~\ref{tab1} and in Figs.~\ref{fig4}, \ref{fig5}, \ref{fig6}.
Results for Argonne18, Paris and NijmII are close reflecting closeness of their wave functions for  $q\lesssim 5$~Fm$^{-1}$. NijmI and CD-Bonn give more distinct results.
Our calculations show that $G_M$ changes sign for all potentials at rather low  $Q$ that is not seen experimentally.
Nevertheless CD-Bonn and NijmI give good results for $G_M$ for $Q<7$~Fm.
Moscow potentials give the best description of charge form factor $G_C$.
\begin{table}[ht]
\caption{Static deuteron form factors.
Two values through slash are relativistic calculation/nonrelativistic calculation.}
\label{tab1}
  \begin{center}
\begin{tabular}{ccc}
\hline\noalign{\smallskip}
 & $G_{M}(0)=\frac{M_{d}}{m_{p}}\mu_{d}$ & $G_{Q}(0)=M_{d}^{2}Q_{d}$  \\
\noalign{\smallskip}\hline\noalign{\smallskip}
Exp & 1.7148 & 25.83 \\
NijmI & 1.697/1.695 & 24.8/24.6 \\
NijmII & 1.700/1.695 & 24.7/24.5 \\
Paris & 1.696/1.694 & 25.6/25.2 \\
CD-Bonn & 1.708/1.704 & 24.8/24.4 \\
Argonne18 & 1.696/1.694 & 24.7/24.4 \\
JISP16 & 1.720/1.714 & 26.3/26.1 \\
Moscow06 & 1.711/1.699 & 24.5/24.2 \\
Moscow14 & 1.716/1.700 & 26.0/25.8 \\
\noalign{\smallskip}\hline
\end{tabular}
\end{center}
\end{table}
The essential factor that influences our calculations  is the nucleon FF dependencies on
the  momentum transferred to  the individual nucleon  $Q_p^2 \approx Q_n^2 \neq Q^2$.
 These FFs are measured at discrete values of $Q_{i=p,n}^2$ while we need dependencies on $Q_i$. In our calculations we take  phenomenological nucleon FF dependencies on $Q_i^2 $ from \cite{FormFBBA}.
It should be noted that neutron EM FFs are extracted from experimental data of $^{2}{\vec {\rm H}} ({\vec e},e'n)p$  and other processes with deuteron and triton using various
  models of these possesses and nuclei. Therefore these FFs are model dependent.
%
We see good general correspondence of the theory and
experiment for $Q< 5$ Fm$^{-1}$. Discrepancies for larger $Q$ are comparable with differences of results for different potentials.
Model calculations  \cite{arenhovel_excange_curr} show that meson exchange currents may give significant effect in EM processes with  $np$-system.
We do not take into account these currents. However it is not clear how these currents may be agreed with short range part of the $NN$ interaction of quantum chromodynamic origin.
Besides the EM FFs of nucleons are not described by meson degrees of freedom at intermediate and high energies  \cite{Perdrisat2007}.

To complete this line of our investigation, we plan to
calculate neutron EM FFs compatible with Moscow potential model which has not been used for the extraction of these FFs.


\begin{thebibliography}{103}
%
\bibitem{Akhiezer}  A. I. Akhiezer, A. G. Sitenko and V. K. Tartakovskii, \textit{Nuclear Electrodynamics} (Springer Series in Nuclear and Particle Physics, 1994).
%
\bibitem{GilmanGross} R. Gilman and Franz Gross, J. Phys. G {\bfseries 28}, R37-R116 (2002).
%
\bibitem{Garson} M. Gar\c{c}on and J. W. Van Orden, Adv. Nucl. Phys. {\bfseries 26},  293 (2001).
%
\bibitem{ArenRitz}
H. Arenh\"{o}vel, F. Ritz and T. Wilbois, Phys. Rev. C {\bfseries 61}, 034002 (2000).
%
\bibitem{AdamAren}
J. Adam, Jr. and H. Arenh\"{o}vel, Nucl. Phys. A {\bfseries 614}, 289 (1997).
%
\bibitem{Tamura}
K. Tamura, T. Niwa, T. Sato and H. Ohtsubo, Nucl. Phys. A {\bfseries 536}, 597 (1992).
%
  \bibitem{AllenKlinkFormF} T. W. Allen,  W. H. Klink and W. N. Polyzou, Phys. Rev. C {\bfseries 63},  034002 (2001).
%
%
\bibitem{LevPace}
F. M. Lev, E. Pace and G. Salm\`e, Phys. Rev. C {\bf 62}  064004 (2000); Nucl. Phys.  A {\bfseries 663}, 365c (2000).
%
%
\bibitem{RKM_Review} B. D.  Keister and W.  Polyzou,
Adv. Nucl. Phys.
 {\bfseries 20},  325 (1991).
 %
\bibitem{Dirac}
   P. A. M. Dirac,  Rev. Mod. Phys. {\bfseries  21},   392 (1949).
     %
\bibitem{BakamjianThomas}  B. Bakamjian and L. H. Thomas, Phys. Rev. {\bfseries 92},  1300 (1953).
%
\bibitem{Sokolov1984}   S. N. Sokolov and A. M. Shatny, Theor. Math. Phys. {\bfseries 37},  1029 (1978).
%
\bibitem{Klink}   W. H. Klink, Phys. Rev. C
{\bfseries 58},   3587 (1998).
%
\bibitem{Desplanques}   B. Desplanques and   L. Theu{\ss}l, Eur. Phys. J. A  {\bfseries  13},   461 (2002).
  %
  %
\bibitem{Lev3}    F. M. Lev,
 Ann.  Phys.  (N. Y.)   {\bfseries  237},   355 (1995); hep-ph/9403222.
 %
 \bibitem{Melde}    T. Melde,  L.  Canton,    W. Plessas and  R. F. Wagenbrunn,  Eur.  Phys.  J.  A  {\bfseries 25},   97 (2005).
     %
%
\bibitem{PFSA1}   T. W. Allen and   W. H. Klink,  Phys. Rev. C  {\bfseries  58},   3670 (1998).
%
\bibitem{PFSA2}   F. Coester and   D. O. Riska,  Few-Body Syst.   {\bfseries  25},   29 (1998).
%
%
\bibitem{Wagen}   R. F. Wagenbrunn,   S. Boffi,   W. Klink, W. Plessas and M. Radici,  Phys. Lett. B {\bfseries 511}, 33 (2001).
%
\bibitem{Amghar}  A. Amghar,   B. Desplanques and   L. Theu{\ss}l,
 Phys. Lett. B {\bfseries 574}, 201 (2003).
%
\bibitem{CoesterFF2}   F. Coester and  D. O.  Riska, Nucl. Phys. A  {\bfseries 728}, 439 (2003).
%
\bibitem{NNscatMy3}
 N. A. Khokhlov and   V. A. Knyr,  Phys. Rev. C {\bfseries 73}, 024004 (2006).
%
\bibitem{ppgammaMy22}
  N. A. Khokhlov,   V. A. Knyr and   V. G. Neudatchin,  Phys. Rev. C {\bfseries 68}, 054002 (2003).
%
\bibitem{gammaDMy1}   V. A. Knyr,   V. G. Neudatchin and    N. A. Khokhlov, Phys. Atom. Nucl. {\bfseries 70}, 879 (2007).
%
%
\bibitem{gammaDMy2}   N. A. Khokhlov,  V. A. Knyr and  V. G. Neudatchin, Phys. Rev. C {\bfseries 75}, 064001 (2007).
%
%
\bibitem{gammaDMy3}   V. A. Knyr,   V. G. Neudatchin and    N. A. Khokhlov, Phys. Atom. Nucl. {\bfseries 70}, 2152 (2007);
 V. A. Knyr and     N. A. Khokhlov, Phys. Atom. Nucl. {\bfseries 66}, 1994 (2008).
%
\bibitem{ednpMy} V. A. Knyr and  N. A. Khokhlov, Phys. Atom. Nucl. {\bfseries 70}, 2066 (2007).
%
%
\bibitem{Arnold81} R. G. Arnold, C. E. Carlson and F. Gross, Phys. Rev. C {\bfseries 23}, 363 (1981).
%
\bibitem{Donnelly} T. W. Donnelly and A. S. Raskin, Ann. Phys. (N.Y.) {\bfseries 169},  247 (1986).
\bibitem{Nijm} V. G. J. Stoks, R. A. M. Klomp, C. P. F. Terheggen and J. J. de Swart, Phys. Rev. C {\bfseries 49}, 2950 (1994).
%
\bibitem{JISP16} A. M. Shirokov, J. P. Vary, A. I. Mazur and T. A.Weber,
Phys. Lett. B {\bfseries 644}, 33 (2007).
%
\bibitem{CD-Bonn} R. Machleidt, Phys. Rev. C {\bfseries 63}, 024001 (2001).
%
\bibitem{Paris} M. Lacombe, B. Loiseau, R. Vinh Mau, J. C\^{o}t\'{e}, P. Pir\'{e}s and
R. de Tourreil, Phys. Lett. B {\bfseries 101}, 139 (1981).

%
\bibitem{Argonne18} R.B. Wiringa, V.G.J. Stoks and R. Schiavilla, Phys. Rev. C {\bfseries 51}, 38 (1995).
\bibitem{Argonne18ms} S. Veerasamy and W. N. Polyzou, Phys. Rev. C {\bfseries 84}, 034003 (2011).
 \bibitem{r87} J. E. Elias, J. I. Friedman, G. C. Hartmann, H. W. Kendall, P.~N.~Kirk, M.~R.~Sogard, L.~P.~Van~Speybroeck and J. K. De Pagter, Phys. Rev. {\bfseries 177},  2075 (1969).
 %
 \bibitem{r90} R. G. Arnold, B. T. Chertok, E. B. Dally, A. Grigorian, C.~L.~Jordan, W.~P.~Sch\"{u}tz, R.~Zdarko, F. Martin and B. A. Mecking, Phys. Rev. Lett. {\bfseries 35},  776 (1975).
 %
\bibitem{r94} R. Cramer \textit{et al.}, Z. Phys. C {\bfseries 29},  513 (1985).
%
 \bibitem{r97} S. Platchkov, A. Amroun, S. Auffret, J.M. Cavedon, P. Dreux, J.~Duclos, B.~Frois, D.~Goutte, H. Hachemi, J. Martino and X. H. Phan, Nucl. Phys. A {\bfseries 510}, 740 (1990).
 %
  \bibitem{r98} L. C. Alexa \textit{et al.}, Phys. Rev. Lett. {\bfseries 82}, 1374 (1999).
  %
 \bibitem{r99} D. Abbott \textit{ et al.}, Phys. Rev. Lett. {\bfseries 82}, 1379 (1999).
 %
  \bibitem{r95} S. Auffret \textit{ et al.}, 
  Phys. Rev. Lett. {\bfseries 54}, 649 (1985).
  %
 \bibitem{r96} P. E. Bosted
 \textit{et al.}, Phys. Rev. C {\bfseries 42}, 38 (1990).
 %
 \bibitem{r76} M. E. Schulze
 \textit{et al.}, Phys. Rev. Lett. {\bfseries 52}, 597 (1984).
 %
 \bibitem{r77} M. Gar\c{c}on
 \textit{et al.}, Phys. Rev. C {\bfseries 49}, 2516 (1994);
 I. The
 \textit{et al.}, Phys. Rev. Lett. {\bfseries 67}, 173 (1991).
 %
 \bibitem{r103} D. Abbott
 \textit{et al.}, Phys. Rev. Lett. {\bfseries 84},  5053 (2000).
 %
 \bibitem{r104} V. F. Dmitriev
 \textit{et al.},
 Phys. Lett. B {\bfseries 157}, 143 (1985).
 %
 \bibitem{r105}   B. B. Voitsekhovskii,   D. M. Nikolenko,  K. T. Ospanov,  S. G. Popov, I.~A.~Rachek, D. K. Toporkov, E. P. Tsentalovich and Yu. M. Shatunov,
 JETP Lett. {\bfseries 43}, 733 (1986).
 %
 \bibitem{r106} R. Gilman
 \textit{et al.}, Phys. Rev. Lett. {\bfseries 65}, 1733 (1990).
 %
 \bibitem{r107} M. Ferro-Luzzi
 \textit{et al.}, Phys. Rev. Lett. {\bfseries 77},  2630 (1996).
 %
 \bibitem{r108} M. Bouwhuis
 \textit{et al.}, Phys. Rev. Lett. {\bfseries 82}, 3755 (1999).
  %
\bibitem{r78} D. Abbott
\textit{et al.}, Eur. Phys. J. A {\bfseries 7}, 421 (2000).
 %
%
\bibitem{FormFBBA} R. Bradford, A. Bodek, H. Budd and J. Arrington, Nucl. Phys. Proc. Suppl. B {\bfseries 159}, 127 (2006).
 %
 \bibitem{arenhovel_excange_curr} H. Arenhoevel, E. M. Darwish, A. Fix and M. Schwamb, Mod. Phys. Lett. A {\bfseries 18}, 190 (2003). %
 %
 \bibitem{Perdrisat2007} C. F. Perdrisat, V. Punjabi and M. Vanderhaeghen, Prog. Part. Nucl. Phys. {\bfseries 59},  694 (2007).

\end{thebibliography}
\end{document}